# Towards the ultimate storage ring: the lattice design for Beijing Advanced Proton Source

XU Gang (徐刚), JIAO Yi (焦毅)
Institute of High Energy Physics, CAS, Beijing 100049, P.R. China

**Abstract:** A storage ring-based light source, Beijing Advanced Proton Source (BAPS) is proposed to store 5-GeV low-emittance electron beam and to provide high-brilliance coherent radiation. In this paper, we report our efforts of pushing down the emittance of BAPS to approach the so-called ultimate storage ring, while fixing the circumference to about 1200 m. To help dealing with the challenge of beam dynamics associated with the intrinsic very strong nonlinearities in an ultralow-emittance ring, a combination of several progressive technologies is used in the linear optics design and nonlinear optimization, such as modified theoretical minimum emittance cell with small-aperture magnets, quasi-3rd-order achromat, theoretical analyzer based on Lie Algebra and Hamiltonian analysis, multi-objective genetic algorithm, and frequency map analysis. These technologies enable us to obtain satisfactory beam dynamics in one lattice design with natural emittance of 75 pm.

## 1 Introduction

A storage ring-based light source, Beijing Advanced Proton Source (BAPS) is planned to be built in Beijing to satisfy the increasing requirements of high-brilliance coherent radiation from the user community. A baseline of BAPS was designed to provide 5-GeV electron beam with natural emittance of 1.6 nanometers (nm) by using 48 double bend achromat (DBA) cells, and of 0.5 nm by using additional damping wigglers [1], within a circumference of 1209 m. Recently we continuously push down the emittance of BPAS in order to approach the so-called ultimate storage ring, i.e., a storage ring with emittance in both transverse planes at the diffraction limit for the range of x-ray wavelengths of interest for scientific community. For BAPS with electron beam of 5 GeV, it requires to reduce the emittance to several tens of picometers (pm).

As a result of the equilibrium between the radiation damping and quantum fluctuation, the natural emittance of electron beam in a storage ring $\varepsilon_{x0}$ is given by [2]

$$\varepsilon_{x0} = C_q \gamma_L^2 \frac{<H>_{\text{dipole}}}{J_x \rho}, \tag{1}$$

where $C_q$ = 3.83*10$^{-13}$ m; $\gamma_L$ is the Lorenz factor; $\rho = L_B/\theta$ is the bending radius of a single dipole, with $L_B$ being the dipole length and $\theta$ the bending angle; $J_x$ is the horizontal damping partition number; and $<H>_{\text{dipole}}$ is the average over the storage ring dipoles of the function

$$H = \gamma_x D_x^2 + 2\alpha_x D_x D_x' + \beta_x D_x'^2, \tag{2}$$

where $\alpha_x$, $\beta_x$, and $\gamma_x$ are the horizontal Courant-Snyder parameters, $D_x$ and $D'_x$ are the horizontal dispersion and its derivative.

In a storage ring with uniform dipoles, $J_x \approx 1$, the theoretical minimum emittance (TME) is derived by minimizing $<H>_{\text{dipole}}$ with symmetric dispersion in the dipole [3],

$$\varepsilon_{x0}^* = \frac{C_q \gamma_L^2 \theta^3}{12\sqrt{15} J_x}, \tag{3}$$

with the optical functions at the dipole center (expressed with subscript of zero) satisfying

$$\beta_{x0}^{*} = \frac{L_B}{2\sqrt{15}}, \quad \alpha_{x0}^{*} = 0, \quad D_{x0}^{*} = \frac{L_B^2}{24\rho}, \quad D_{x0}'^{*} = 0. \tag{4}$$

In Eqs. (3) and (4), an asterisked quantity means the quantity is evaluated when the exact TME condition is fulfilled.

In this paper, a TME cell refers to a unit cell satisfying the conditions in Eq. (4) and having horizontal phase advance of half-cell $\mu_x^*$ = 142 degrees. It consists of one dipole and symmetric quadrupole structure outside, with focusing quadrupole (QF) closer to the dipole (or with focusing gradient combined into the dipole). A TME-like cell refers to a unit cell with similar layout and with optical functions at the dipole center close to, but not exactly on the conditions in Eq. (4). In Ref. [4], one of the authors (JIAO Y) highlights one kind of TME-like cells, named the modified-TME cell, in which the defocusing quadrupole (QD) is closer to the dipole or the defocusing gradient is directly combined into the dipole. Figure 1 presents the configurations of a modified-TME cell and a conventional TME-like cell (or a TME cell). Study shows that such kind of TME-like cell allows minimal emittance of the order of about 3 times of the theoretical minimum, phase advance per half cell below $\pi/2$, relaxed optical functions, moderate natural chromaticities, and most importantly, a compact layout [4].

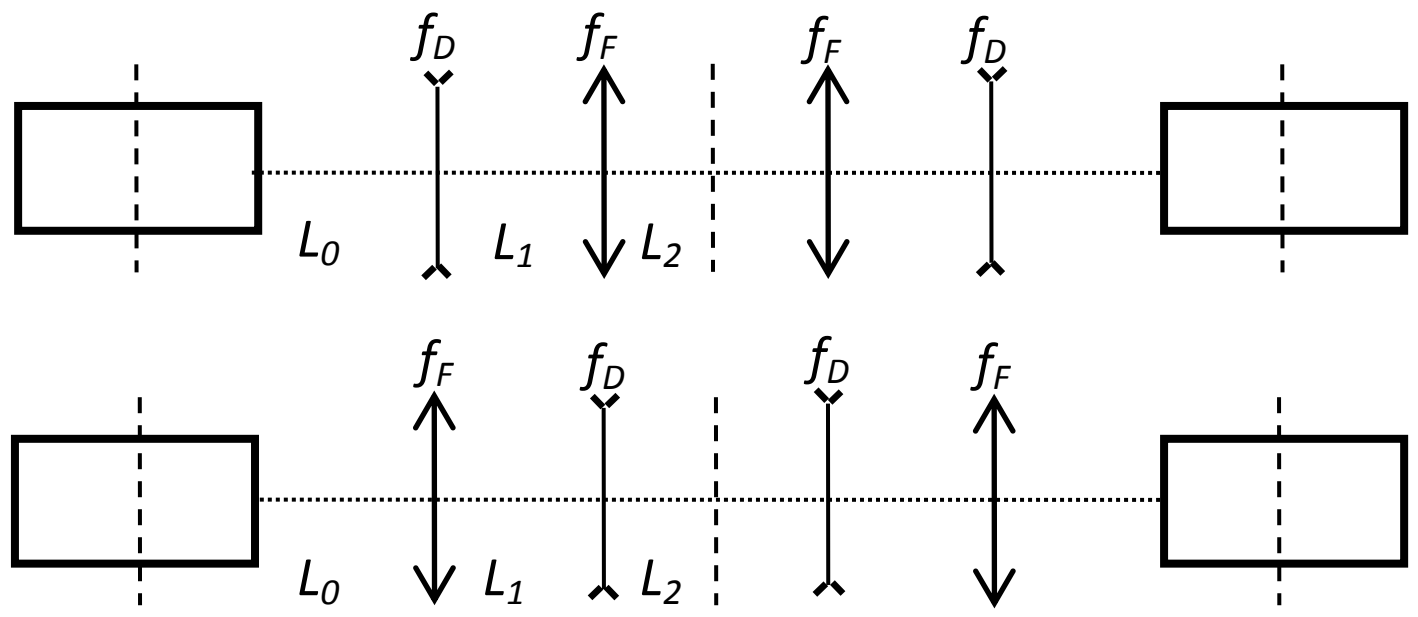

Fig. 1. Two TME-like cell configurations, with QD closer to the dipole in a modified-TME cell (the upper plot), and with QF closer to the dipole in a conventional TME-like cell or in a TME cell (the lower plot).

Because of the ability of approaching emittance of the theoretical minimum, TME or TME-like cell has been adopted as the basic unit cell in many ultralow emittance designs [5-9]. From Eq. (3), the TME is proportional to $\theta^3$. To reduce the natural emittance of an electron storage ring composed of TME or TME-like cells and with fixed beam energy and circumference, the most effective way is to reduce the bending angle of the dipole $\theta$, which means increasing the number of the dipoles $N_d$ as well as the number of the cells $N_c = N_d$ in a ring associated with decreasing cell length. To reach a small cell length, and at the same time, to achieve relatively low natural chromaticities, we use the modified-TME cells with combined-function dipoles and small-aperture magnets in our design.

According to a rough scaling that $S \propto \varepsilon_x^{-2/3}$ with $S$ the integral strength of the chromaticity-correction sextupoles [4], decreasing the emittance in an electron storage ring unavoidably leads to an increase of the required sextupole strength. In an ultralow emittance design, the required sextupole strength can be so strong that the nonlinear dynamics associated with the very large nonlinear geometric and chromatic aberrations from the sextupoles becomes a great challenge to the performance of the storage ring. To preserve large enough injection acceptances as pushing down the emittance to several tens of pm region, more advanced optimization technologies or methods compared to those used for the present 3rd generation rings, are essentially required. In our design, we construct a quasi-3rd-order achromat within every eight superperiods through fine tuning of the phase advance, so as to approximately cancel most of the 3rd- and 4th-order resonances [9-11]; we use multi-families of sextupoles and octupoles to minimize the residual 3rd- and 4th-order resonance driving terms, and to control other nonlinear terms to an acceptable level. Using a theoretical analyzer based on Lie Algebra and Hamiltonian analysis, we obtain expressions of the nonlinear terms with respect to the sextupole and octupole strengths; we then set three objectives characterizing the chromaticity, detune and resonance terms and search for a set of optimal solutions with non-dominated sorting genetic algorithm II (NSGA-II [12]); after numerical tracking with AT code [13] and frequency map analysis [14], we can find the best solution that promises large enough dynamic aperture (DA).

In the following we will present one lattice design for BAPS reaching natural emittance of 75 pm. Linear optics is shown in Sec. II, where we focus on the design of the modified-TME unit cell and the quasi-3rd-order achromat. Nonlinear dynamics is discussed in Sec. III, with an emphasis on the extraction of the chromaticity, detune and resonance terms with the theoretical analyzer and subsequent optimization with NSGA-II. Conclusions are given in Sec. IV.

## 2 Linear optics design

As mentioned above, as we pursue an ultralow emittance ring composed of TME-like cells, the number of dipoles $N_d$ and hence the number of cells $N_c$ increase as the emittance decreases. Limited by the construction budget, the BAPS circumference is expected to be around 1200 m. As a balance of the requirements of larger $N_c$ for lower emittance and of more straight sections for insertion devices (ID), we adopt 32 supercells consisting of seven-bend-achromat (7BA). Each supercell includes five modified-TME unit cells and two dispersion matching cells at two ends to produce dispersion-free straight section, as shown in Fig. 2. The bending angle of the outer dipole in a supercell is set to 1/2 (not far from the exact value $3^{-1/3}$ [15]) of the bending angle of the inner dipoles to obtain a minimum emittance. Overall there are equivalently 192 dipoles in the ring. From Eq. (3), the theoretical minimum emittance $\varepsilon_x^*$ = 27.7 pm with $J_x$ = 1.

As a simple but reasonable assumption, we presume that the matching cell length ($L_{dm}$) is 1.5 times the modified-TME unit cell length ($L_{mt}$), i.e. $L_{dm}$ = 1.5 $L_{mt}$. With an additional assumption that the straight section is 8 m in each supercell, we obtain a linear relation between the circumference and the modified-TME unit cell length, as shown in Fig. 3. The circumference varies significantly, from 1024 to 1408 m, while changing $L_{mt}$ from 3 to 4.5 m.

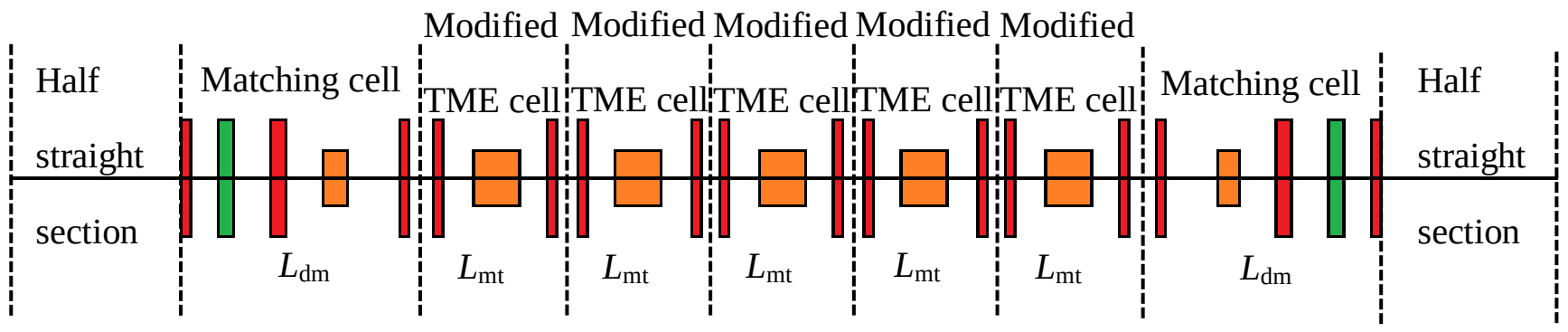

Fig.2. Schematic of one 7BA supercell.

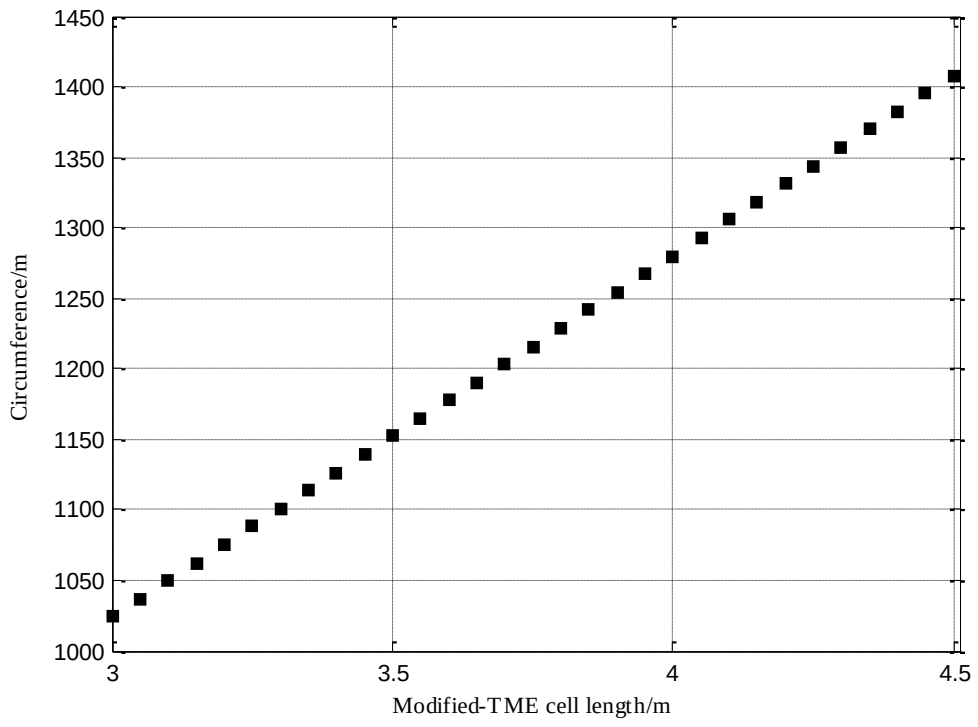

Fig. 3. Estimation of the circumference of the BAPS ring with varying unit cell length.

To control the circumference and hence the total cost, special attention should be paid to the shortening of the modified-TME unit cell length. For this purpose, we follow the MAX IV magnetic design [16], i.e. combination of the defocusing gradient in the dipole (resulting in $J_x > 1$ for even lower emittance), adoption of small-aperture (25 mm diameter) magnets, and integration of the magnets into a common iron block. Small magnet aperture allows high available strength of the quadrupole and sextupoles (up to 2.8 $m^{-2}$ and 460 $m^{-3}$, respectively) and hence allows short quadrupole and sextupoles (as low as 0.25 m), provided the available maximum magnetic pole face field is 0.6 Tesla. On the other hand, we use a relatively long combined-function dipole (1.2 m), in order to provide relaxed horizontal beta and dispersion functions [see Eq. (4)] and hence to release the required chromaticity-correction sextupole strength. We also reserve a 0.8 m space in each unit cell to accommodate diagnostics and correctors. Finally we reach a cell length of 3.8 m. With such a compact layout, the length of each element has little variable range, and only the QF strength $K_f$ and the defocusing gradient combined in the dipole $K_d$ can be varied. We can therefore derive the expressions of the emittance [according to Eq. (1)] and the phase advance with respect to $K_f$ and $K_d$, and view the variations of these parameters in ($K_f$, $K_d$) space, as shown in Fig. 4. From the figure, one can clearly know the available range of $K_f$ and $K_d$ for stable optics, easily match the phase advance to an expected value, and meanwhile find out the corresponding emittance. In our design we choose ($K_f$, $K_d$) = (2.619, 0.692) to reach phase advance per half cell ($\mu_x$, $\mu_y$) = (13$\pi$/32, 7$\pi$/64), natural chromaticity per cell ($\xi_x$, $\xi_y$) = (-0.75, -0.36), and natural emittance of 80 pm (~ 3 times the theoretical minimum). The optical functions in a modified-TME unit cell are shown in Fig. 5.

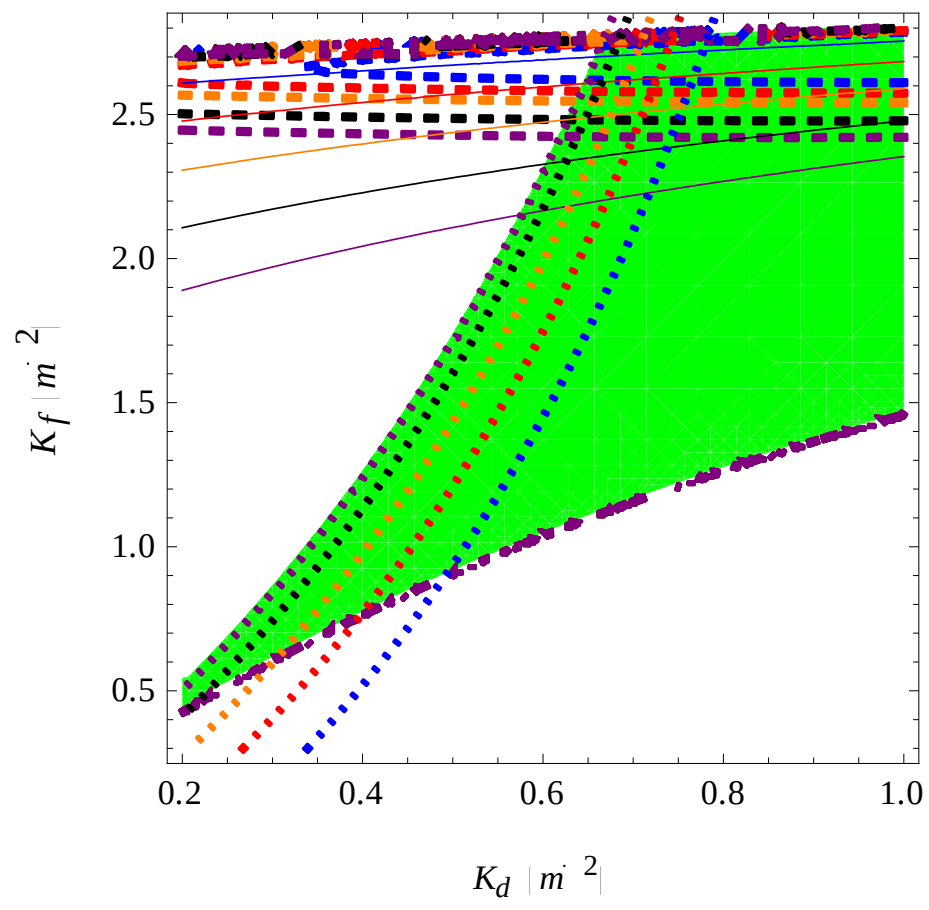

Fig.4. (color online) Necktie diagram and contour plots of the emittance and the phase advance in ($K_f$, $K_d$) space. The green area represents the region for stable optics. The solid lines, from upper to lower, refer to $\mu_x$ = (14/32, 13/32, 12/32, 11/32, 10/32) $\pi$; the dotted lines, from left to right, refer to $\mu_y$ = (3/64, 5/64, 7/64, 9/64, 11/64) $\pi$; and the dashed lines, from upper to lower, refer to $\varepsilon_x$ = (80, 85, 90, 100, 110) pm.

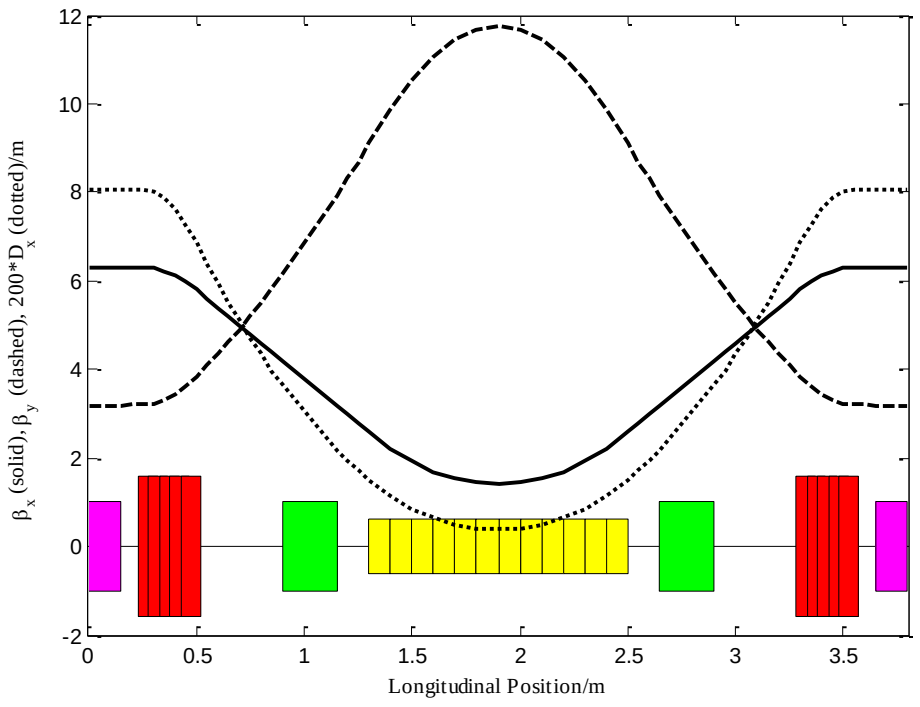

Fig. 5. Optical functions in a modified-TME unit cell with ($\mu_x$, $\mu_y$) = (13$\pi$/32, 7$\pi$/64).

The linear optics design on one hand should allow low beta functions in the straight section for high brilliance (optimal beta $\beta = L_{ID}/2\pi$ [9], with $L_{ID}$ the ID length), and on the other hand, should promise large enough acceptance for off-axis injection. Like that at NSLS-II [17], we merge two 7BA supercells into one superperiod. Each superperiod has a high-beta 10-m straight section for injection and a low-beta 6-m straight section for IDs. The phase advance of each superperiod is chosen to be $12\pi + \pi/4 + \delta\upsilon_x*\pi/8$ in horizontal plane and $4\pi + \pi/4 + \delta\upsilon_y*\pi/8$ in vertical plane, where $\delta\upsilon_x$ and $\delta\upsilon_y$ are the expected decimal portions of the working point of the ring. Figure 6 shows the optical functions in a superperiod. Note that the third term for the phase advance is much smaller than the sum of the first two. Thus, every eight of superperiods makes an approximate identity transformation and forms a quasi-3rd-order achromat. Such a design helps to approximately cancel the 3rd- and most of the 4th- order resonances and hence to facilitate the subsequent nonlinear optimization (see Sec. III for more discussions).

Sixteen identical superperiods compose a 1263.4-m ring with natural emittance of 75 pm. The main parameters of the ring are summarized in Table I.

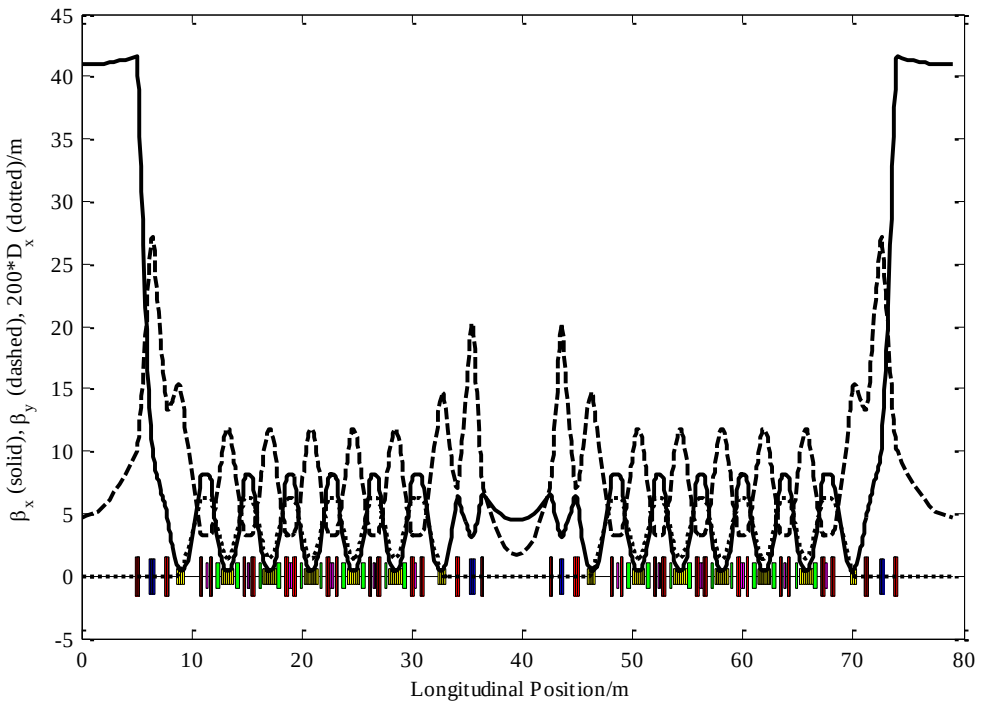

Fig. 6. Optical functions in a superperiod consisting of two 7BA supercells.

**3 Nonliear optimization**

The aim of the nonlinear optimization is to obtain large enough DA and momentum acceptance for off-axis injection and a good performance of the ring. As mentioned above, the sextupole strength increases as the emittance decreases. In spite of the adoption of the modified-TME cell with relatively low phase advance and relatively long combined-function dipoles, the required sextupole strengths to compensate the natural chromaticity are still very large, i.e., $K_{sf}$ = 290 $m^{-3}$ and $K_{sd}$ = –274 $m^{-3}$, provided only two families of sextupoles (SF and SD) are used for chromaticity correction. Strong nonlinearities induced by the sextupoles may limit DA to a few millimeters (mm) or even lower. Compared to the nonlinear optimization for the 3rd generation rings, more advance tools or methods are required to analyze and control the nonlinear terms to obtain a satisfactory beam dynamics.

As a start of the optimization, and also for demonstration of the effect of quasi-3rd-order achromat, we first perform numerical tracking (without magnetic erros) and FMA for the case with only two families of chromaticity-correction sextupoles. The results are shown in Fig. 7. The on-momentum horizontal DA is 7.5 mm, larger than the requirement (≥ 4 mm) of off-axis injection with pulsed sextupoles [18]. The resonances have small driving terms (verified by the following analysis) due to approximate cancellation, and hence do not cause significant distortions in the frequency map (FM). In the absence of magnetic errors, the particles can pass through the integer resonances at x = 4.5 mm and y = 1.5 mm without loss. However, the rule of thumb is that the integer resonances are always dangerous in a realistic machine and can-not be passed. One can foresee that when the integer resonances are more excited due to magnetic field errors and misalignments, all the orbits beyond the integer resonances will become unstable, leading to a significant shrinkage in DA.

Table I: Main parameters for the BAPS 5 GeV storage ring

| **Parameters** | **Flux** | **Unit** |
|---|---|---|
| Energy | 5 | GeV |
| Circumference | 1263.4 | m |
| Horizontal damping partition number $J_x$ | 1.40 | |
| Natural emittance | 75 | pm |
| Working point (H/V) | 98.4/34.3 | |
| Natural chromaticities (H/V) | -189/-113 | |
| Number of 7BA achromats | 32 | |
| Number of high-beat 10-m straight sections | 16 | |
| Beta functions in high-beta straight section (H/V) | 41/4.7 | m |
| Number of low-beat 6-m straight sections | 16 | |
| Beta functions in low-beta straight section (H/V) | 4.5/1.7 | m |
| Damping times (x/y/z) | 20/28/17.4 | ms |
| Energy spread | $8\times10^{-4}$ | |
| Momentum compaction | $3.86\times10^{-5}$ | |

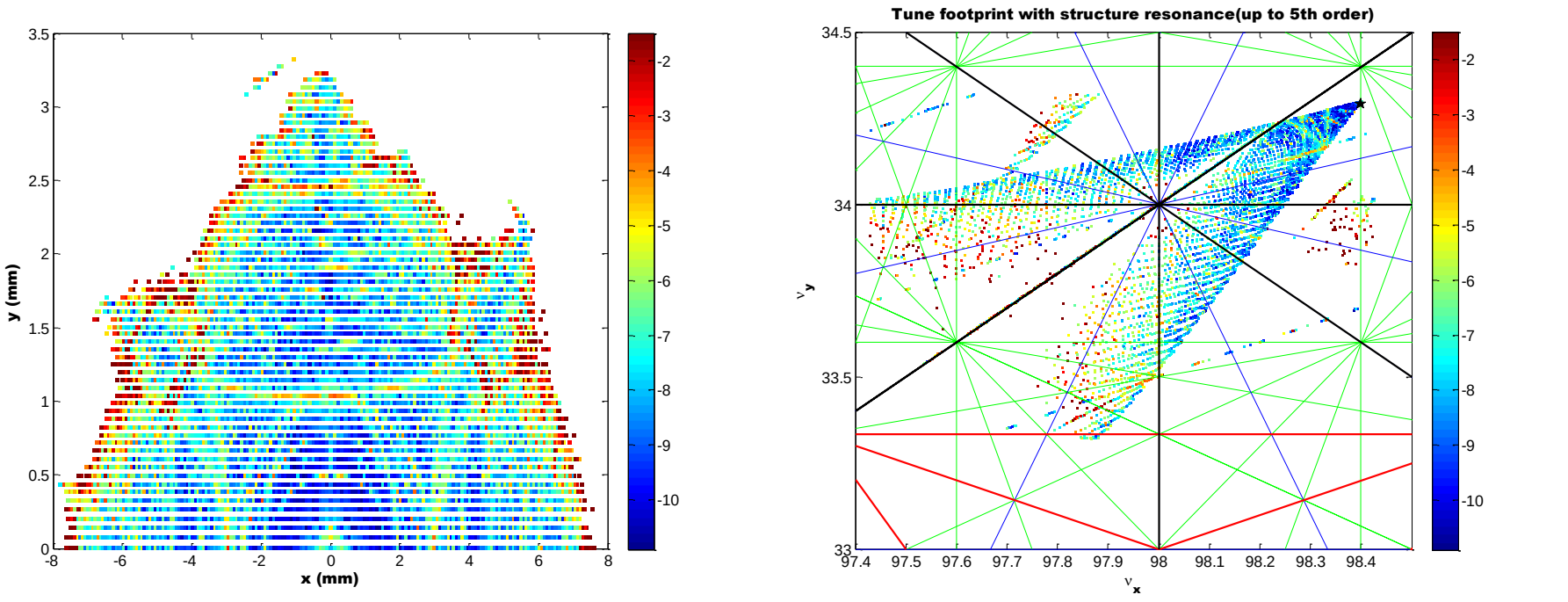

Fig. 7. (color online) The dynamic aperture and frequency map obtained after tracking of 1024 turns with AT code for BAPS ring lattice with only two families of chromaticity-correction sextupoles. The colors, from blue to red, represent the stabilities of the particle motion, from stable to unstable.

Further study shows that the large detune terms are responsible for particles quickly reaching the integer resonances. In order to minimize the detune terms, and at the same time, to control the other nonlinear terms to an acceptable level, we use additional four families of chromaticity-correction sextupoles and six families of harmonic sextupoles and octupoles for nonlinear optimization. To understand the combined effects of multi-families of sextupoles and octupoles, one of the authors (XU G) develops a theoretical analyzer based on Lie Algebra and Hamiltonian dynamics (see Appendix for a short introduction), from which we can obtain analytical expressions of the detune (up to the 2nd order), chromaticity (up to the 4th order), and the resonance driving terms (up to the 6th order) with respect to the sextupole and octupole strengths. It is worth mentioning here that the resonance driving terms obtained by the analyzer are somewhat different from those obtained with normal form method [19], which, however, does not affects the effectiveness of the analyzer in measuring the resonance strengths. We then make multi-objective genetic optimization with NSGA-II by setting three objective functions $f_1$, $f_2$, and $f_3$ to characterize the detune, chromaticity and resonance terms, respectively. Figure 8 presents the Pareto-optimal solutions obtained after 500 generations with NSGA-II. Among the obtained optimal solutions, we select those providing good balance of three objectives and verify them with numerical tracking and FMA. Finally, we obtain one optimal set of the sextupole and octupole strengths. The nonlinear terms for this solution are listed in Table II and compared with those for the case with only two families of chromaticity-correction sextupoles. One can see that after optimization the detune terms are smaller, while the chromatity and resonance terms do not increase a lot. The corresponding on-momentum DA and FM are shown in Fig. 9. In this case, particles reach the integer resonance at relatively large amplitudes, i.e., x = 6 mm and y = 2.6 mm, which promises off-axis injection in the high-beta 10-m straight section.

Table II: Nonlinear terms with only two families of sextupole (TFS) or multi-families of sextupoles and octupoles (MFSO) for BAPS ring

| **Parameters** | **TFS** | **MFSO** |
|---|---|---|
| First order detune ($dQ_x/dJ_x$, $dQ_x/dJ_y$, $dQ_y/dJ_y$) | (9.6, 3.8, 7.2) $\times 10^6$ | (5.8, 0.6, 0.6) $\times 10^6$ |
| Second order detune ($dQ_x^2/dJ_x^2$, $dQ_y^2/dJ_y^2$) | (1.1, 1.3) $\times 10^{12}$ | (3.2, 3.8) $\times 10^{12}$ |
| Horizontal chromaticities ($\xi_x$, $\xi_x'$, $\xi_x''$, $\xi_x'''$) | (0, -1271, -7874, $1.5\times 10^7$) | (0, -769, $1.2\times 10^5$, $4\times 10^7$) |
| Vertical chromaticities ($\xi_y$, $\xi_y'$, $\xi_y''$, $\xi_y'''$) | (0, -442, $-2.8\times 10^5$, $1.1\times 10^7$) | (0, -755, $-2.2\times 10^5$, $9\times 10^6$) |
| Sum of resonance driving terms (1st, 2nd, 3rd, 4th, 5th, 6th) | ($3\times 10^4$, $1\times 10^{10}$, $3.8\times 10^5$, $1\times 10^{12}$, $7\times 10^{16}$, $6\times 10^{23}$) | ($3\times 10^4$, $7\times 10^{10}$, $3.9\times 10^5$, $2.5\times 10^{12}$, $1\times 10^{16}$, $1\times 10^{26}$) |

Further study shows that this optimal solution allows momentum acceptance of ~ 3%, which ensures long enough Touschek lifetime, up to 5 hours. The emittance can be further reduced to 16 pm by installing totally 60-m superconducting damping wigglers (3.3 Tesla, 3.1 cm period) in the straight sections. Another option is to increase the circumference to 1500 m. Studies show that the natural emittance can be reduced to 30 ~ 40 pm. In this case, the required total length of the damping wigglers to achieve emittance of 10-pm level is expected to be smaller. One interesting and important topic related to the design an ultralow design is to achieve a round beam, i.e. with equivalent transverse emittances. One of the authors (XU G) has proposed a novel method of producing a locally-round beam by using solenoid and anti-solenoid, which, however, will be discussed in detail elsewhere [20]. With this method, we can achieve electron beam with emttance $\varepsilon_x = \varepsilon_y = 8$ pm in the presence of the damping wigglers, without introducing great perturbations to the global beam dynamics of the ring.

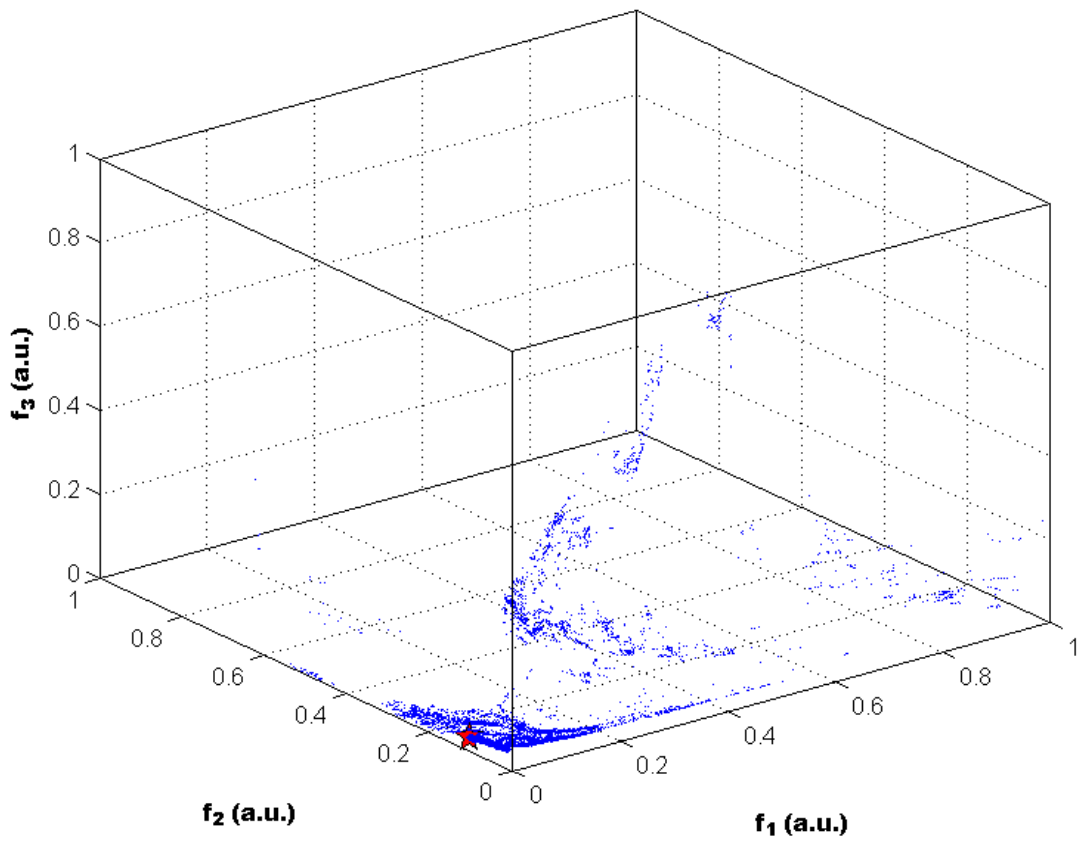

Fig. 8. Pareto-optimal solutions obtained after 500 generations with NSGA-II. Three objectives are used, with $f_1$, $f_2$, and $f_3$ characterizing the detune, chromaticity and resonance terms, respectively. The star denotes the best solution found, which provides large dynamic aperture.

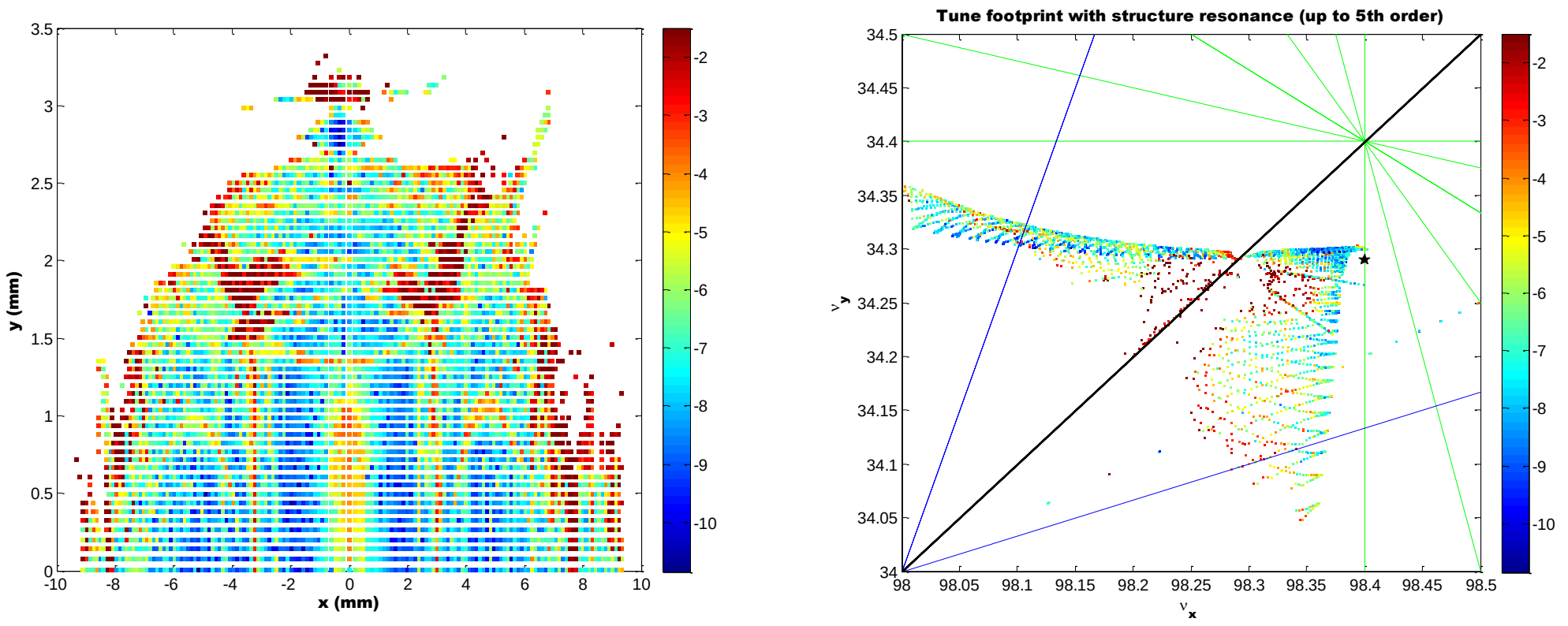

Fig. 9. (color online) The dynamic aperture and frequency map obtained after tracking of 1024 turns with AT code for BAPS ring lattice by using 12 families of sextupoles and 6 families of octupoles with strengths obtained by NSGA-II. The colors, from blue to red, represent the stabilities of the particle motion, from stable to unstable.

## 4 Conlusions

Decreasing the natural emittance of an electron storage ring to several tens of picometers will lead to a great challenge to the linear optics design and nonlinear optimization. In this paper, we present one lattice design for the 5-GeV BAPS storage ring with natural emittance of 75 pm. We adopt the modified-TME cell with small-aperture magnets as the unit cell, so as to realize compact layout and control the required field strengths of the chromaticity-correction sextupoles. We make every eight superperiods as a quasi-3rd-order achromat so as to approximately cancel most of the 3rd- and 4th-order resonances. Using a theoretical analyzer, we obtain the expressions of the nonlinear terms with respect to the sextupole and octupole strengths. The obtained dependency enables us to search for a set of solutions by optimization with NSGA-II. After numerical tracking and frequency map analysis, we find one optimal set of the sextupole and octupole strengths which promises off-axis injection and long enough lifetime. To summarize, the techniques used in the lattice design for BAPS help us deal with the great challenge associated with the intrinsic strong nonlinearities, which would also be beneficial for other ultralow emittance designs. In the end, we have to state that except the sextupoles and octupoles mentioned in this paper, there are other nonlinearity sources in a ring, such as magnetic field error and misalignment, sinusoidal field of the damping wigglers and undulators, whose effects to the beam dynamics should and will be considered in our next work.

**APPENDIX Computation of Analytical Hamiltonian in a Storage Ring**

A theoretical analyzer was developed on Mathematica platform to compute the one-turn Hamiltonian as a function of more than 23 strength variables of sextupole, octupole, decapole and dodecaples. From the one-turn Hamiltonian, one can get the analytical formula of the high order chromaticities, detune and resonance terms.

In the Lie Algebra framework [21], the map for an accelerator element with index $i$ can be represented as exponential operator,

$$X_1 = e^{-:L_i H_i:} X |_{X=X_0} = \sum_{k=0}^{\infty} \frac{(-1)^k L_i^k}{k!} :H_i^k: X |_{X=X_0}, \tag{A1}$$

where X = ($x$, $p_x$, $y$, $p_y$)’, $X_1$ and $X_0$ refer to the initial and final canonical coordinates, $L_i$ is the length of the element, and $H_i$ is the Hamiltonian describing the particle motion in the element, which is in the form of [22]

$$H_i = hx + h^2 x^2/2 - (1+hx)\sqrt{1-2\delta/\beta+\delta^2-p_x^2-p_y^2} - \delta/\beta + \tilde{V}, \tag{A2}$$

with

$$\begin{aligned}\tilde{V} = {} & \frac{k_1}{2}(x^2-y^2) + h_1 xy + k_2\frac{x^3-3xy^2}{6} + h_2\frac{3x^2y-y^3}{6} + k_3\frac{x^4-6x^2y^2+y^4}{24} + h_3\frac{4x^3y-4xy^3}{24} \\ & + k_4\frac{x^5-10x^3y^2+5xy^4}{120} + h_4\frac{5x^4y-10x^2y^3+y^5}{120} \\ & + k_5\frac{x^6-15x^4y^2+15x^2y^4+5xy^4}{720} + h_5\frac{6x^5y-20x^3y^3+6xy^5}{120},\end{aligned}$$

where $h$ = 1/ρ, $\beta$ = $v/c$, $k_i$ and $h_i$ are the regular and skew strength of the multipoles, $\delta$ is momentum deviation relative to the reference momentum and is taken a parameter here.

From Eq. (A1), we obtain the corresponding Taylor map of the element (truncated to 5th power of the canonical variables) through straightforward derivations,

$$X_1 = F_i(X_0), \tag{A3}$$

with the $j$th final coordinate being analytic in $X_0$,

$$X_1(j) = \sum_k R_{jk} X_0(k) + \sum_{kl} R_{jkl} X_0(k) X_0(l) + \ldots + \sum_{klmno} R_{jklmno} X_0(k) X_0(l) X_0(m) X_0(n) X_0(o).$$

In this way, the Taylor maps for all the elements can be obtained and they can be composed to the one-turn Taylor map in a similar form with Eq. (A3). During the calculation, the map is always truncated to 5th power of the variables. Note that because of the truncation, the Taylor map is not symplectic.

Considering the one-turn Hamiltonian has the following form

$$H_{eff} = \sum_{jk} h_{jk} X_0(j) X_0(k) + \ldots + \sum_{jklmno} h_{jklmno} X_0(j) X_0(k) X_0(l) X_0(m) X_0(n) X_0(o). \quad \text{(A4)}$$

From the correspondence of the Lie map and Taylor map, we can find a set of linear equations between the coefficients of the one-turn Hamiltonian and the one-turn Taylor map. Solving the linear equations order by order, one can obtain $h_{jk}$, $h_{\mathrm{jkl}}$, …,$h_{jklmno}$.

We make canonical transformation from the coordinate-momentum variables ($x$, $p_x$, $y$, $p_y$)' to action-angle variables ($J_x$, $\phi$, $J_y$, $\varphi$), using the generating function of

$$F_1(x,\phi,y,\varphi) = -\frac{x^2}{2}\tan(\phi) - \frac{y^2}{2}\tan(\varphi). \quad \text{(A5)}$$

The new Hamiltonian is in the form

$$H_{eff,new} = H_0(J_x, J_y, \delta) + H_1(J_x, J_y, \phi, \varphi) + G(J_x, J_y, \phi, \varphi, \delta). \quad \text{(A6)}$$

From $H_0$ one can derive the analytical expressions of the chromaticity and detune terms by taking derivative of $H_0$ with respect to $\delta$ and $J_{x,y}$, respectively; and taking derivative of $H_1$ one can obtain the expressions of the resonance coefficients.

**Acknowledgement**

This work is supported by the special fund of Chinese Academy of Sciences No. H9293110TA.